\documentstyle[preprint,aps,rotate,psfig]{revtex}
\begin{document}
\draft
\title{
Neutrino absorption cross sections in supernova
environment}
\author{
J.M. Sampaio$^1$, K. Langanke$^1$ and
G. Mart\'{\i}nez-Pinedo$^2$}  

\address{$^1$Institut for Fysik og Astronomi, {\AA}rhus Universitet,
  DK-8000 {\AA}rhus C, Denmark\\
$^2$Departement f\"ur Physik und Astronomie der Universit\"at Basel,
Basel, Switzerland 
}
\date{\today}
\maketitle

\tightenlines

\begin{abstract}
We study charged-current neutrino cross sections on neutronrich nuclei
in the mass $A\sim60$ region. Special attention is paid to environmental
effects, i.e. finite temperature and density, on the cross sections. As
these effects are largest for small neutrino energies, it is sufficient
to study only the Gamow-Teller (GT) contributions to the cross sections. The
relevant GT strength distributions are derived from large-scale
shell model calculations. We find that the low-energy cross sections are
enhanced at finite temperatures. However, for
$(\nu_e,e^-)$ reactions  Pauli blocking of the electrons in the final
state makes the cross sections for low-energy neutrinos
much smaller than for the competing
inelastic scattering on electrons at moderate and large densities.
Absorption cross sections  for low-energy antineutrinos are strongly 
enhanced at finite
temperatures. 
\end{abstract}


\narrowtext

It has long been recognized that neutrino-induced reactions play a
crucial role during the collapse stage of a type II supernova
\cite{Arnett}. 
In particular, elastic neutrino scattering off nuclei reduces the
neutrino mean-free path and leads eventually to  `neutrino trapping'
at densities in excess of a few
$10^{11}$ g/cm$^3$. Subsequently the neutrinos in the core are
thermalized by inelastic neutrino-electron scattering.
Obviously collapse simulations require
a detailed description of neutrino transport which is nowadays achieved
within Boltzmann neutrino transport calculations and should in principle
include all potentially important neutrino reactions
\cite{Bruenn85,Mezzacappa}.  Haxton
pointed out that charged- and neutral-current reactions on nuclei
should be included in these simulations \cite{Haxton88}. Interpreting
$^{56}$Fe as a representative nucleus the relevant neutrino-nucleus
cross sections have been calculated in \cite{Bruenn91} based on the
allowed and first-forbidden ground-state response and compared to other
weak-interaction processes under core-contraction conditions.

Neutrino-induced reactions are usually dominated by capture to the giant
resonances. As a consequence allowed (Fermi and Gamow-Teller)
transitions determine the neutrino cross sections at low energies, while
first-forbidden transitions cannot be neglected for neutrino energies in
excess of say 20 MeV. We also note that for a nucleus with negative
$Q$-value, like $^{56}$Fe, a minimal neutrino energy is required.
Consequently the cross section drops sharply at low energies. However,
for neutronrich nuclei, as typically encountered during the late core
contraction stage, the $Q$-value is positive allowing $(\nu_e,e^-)$
reactions for all neutrino energies. On the contrast, $({\bar
\nu_e},e^+)$ reactions have to overcome the reaction $Q$-value. The
picture described here corresponds to neutrino reactions on the nuclear
ground state. 

Now we will turn our discussion to neutrino reactions on nuclei at
finite temperature, described by a Boltzmann distribution of nuclear
states. There are giant resonances built on all these excited states and
due to Brink's hypothesis the excitation energy of these resonances in
the daughter nucleus are the same as for the ground state, only shifted
by the excitation energy of the parent state \cite{Brink}. 
If Brink's hypothesis is
exactly valid, the nuclear response is independent of the initial state
and the cross section for the nuclear ground state is the same as at
finite temperature. We will demonstrate in the following that this
picture, put forward in \cite{Bruenn91}, is actually valid at low
temperatures. However, at the temperatures relevant for the core
contraction stage ($T>10^{10}$ K) there are states in the thermal
ensemble which do not fulfill the Brink hypothesis. These are the giant
resonances built on the low-lying daughter states in the inverse
reaction. Fuller {\it et al.} \cite{FFN} have noticed the importance of
these states for stellar beta-decay (equivalent to the $(\nu_e,e^-$)
reaction) and have introduced the name `backresonances' for these
states. We will in the following see that, for supernova conditions,
the backresonances are more important for $({\bar \nu_e},e^+)$
reactions.  The backresonances are expected to have, besides the giant
resonance contribution from Brink's hypothesis, an additional component
with a sizable nuclear matrix element. Importantly this component is
particularly favored  by phase space for small neutrino energies. 

From this general discussion, which will be supported by our
calculations, we expect that it is sufficient to study the finite
temperature effects on the Gamow-Teller (GT) contribution to the
neutrino-nucleus cross section as i) this component dominates the cross
section at low neutrino energies and ii) the GT giant resonances reside
at significantly lower excitation energies than those for forbidden
responses, i.e. the GT backresonances are thermally more easily
accessible than backresonances for other multipoles. We note that
there are no Fermi backresonances in $(\nu_e,e^-)$ reactions, while for
$({\bar \nu_e},e^+)$ reactions on neutronrich nuclei these states are at
too high excitation energies in the parent nucleus to be of importance.
For neutrino scattering at low energies it is  valid to neglect the
dependence of the nuclear multipole operators on momentum transfer.

Restricting ourselves to the GT contribution only, the neutrino-nucleus
cross section is given by
\begin{equation}
\sigma (E_\nu) = \frac{G_F^2 {\rm cos}^2 \theta_C}{\pi} 
\sum_{if} \frac{(2J_i+1) exp[-E_i/kT]}{G}
k_e^{if} E_e^{if}
F(\pm Z+1,E_e^{if})  B_{if} (GT)
\end{equation} 
where $G_F$ is the Fermi constant, $\theta_C$ the Cabibbo angle, and
$k_e$ and $E_e$ the momentum and energy of the outgoing electron or
positron. The function $F(Z,E_e)$ corrects for the Coulomb distortion of
the outgoing lepton wave function, where the plus sign refers to
electrons. The sum is over the final ($f$) and
initial ($i$) nuclear states, where the latter have excitation energy
$E_i$ and angular momentum $J_i$, and $G$ is the nuclear partition
function.  
Finally,  $B_{if}(GT)$ defines the $B(GT)$ value 
between the initial and final
nuclear states. 
The nuclear structure information used in this paper
(spectra and $B(GT)$ values) are based on large-scale state-of-the-art
shell model calculations. These studies are in detail described in Refs.
\cite{Caurier99,Langanke00} which also demonstrate that these shell
model studies reproduce the nuclear properties, including the GT
strength distributions, quite well.

To understand the relevance of the various components our investigations
have been performed in a sequence of steps. To demonstrate the results we
have chosen the nucleus $^{56}$Fe, following \cite{Bruenn91}.
At first we evaluate the
cross section (1) by explicitly including the lowest excited states, but
neglecting the contributions of the backresonances. Fig. 1a compares the
$(\nu_e,e^-)$ cross section as calculated for the ground state $(i=0)$
with the 
one obtained
by considering the lowest 4 states in $^{56}$Fe. The study has been
performed for the same condition as adopted in Fig. 1a of
\cite{Bruenn91}, i.e. $T=10^{10}$ K (equivalently 0.86 MeV). 
Supporting the argumentation of
\cite{Bruenn91}  we find nearly
identical results in both approaches, as the Brink hypothesis is well
fulfilled for these lowest states (other examples for the validity of
this hypothesis are given in \cite{Langanke00}).

In the second step we study the influence of the backresonances on the
cross section at finite temperature. We will make the assumption that
the GT strength distribution 
$S_{GT}^i (E_f)$ 
for an initial state $i$
can be split into two components 
$S_{GT}^i (E_f) = S_{Brink}^i (E_f) + S_{back}^i (E_f)$ where the first
component obeys Brink's hypothesis, i.e. $S_{Brink}^0 (E_f) = 
S_{Brink}^i (E_f+E_0)$, and the second component represents the
backresonances built on the lowest states in the daughter nucleus. 
Upon inserting into Eq. (1) the sum over the first component reduces to
the ground state contribution . We then have
\begin{eqnarray}
\sigma (E_\nu)  =  \frac{G_F^2 {\rm cos}^2 \theta_C}{\pi} 
\{
& \sum_f & k_e^{f} E_e^{f}
F(\pm Z+1,E_e^{f})  B_{0f} (GT)  \nonumber \\ 
+ 
& \sum_{if} &\frac{(2J_i+1) exp[-E_i/kT]}{G}
k_e^{if} E_e^{if}
F(\pm Z+1,E_e^{if})  B_{back,if} (GT)
\}
\end{eqnarray} 
The nuclear matrix elements for the backresonance part have been derived
from large-scale shell model calculations \cite{Caurier99} for the lowest 
(typically 4-12) daughter states
in the inverse direction. B(GT) values, which connect these daughter
states to the parent ground state, have been eliminated in the
backresonance contribution to avoid double-counting. The partition
function $G$ has been calculated from the many (several hundred)
backresonances included in our calculations.

Fig. 1b shows the $(\nu_e,e^-)$ cross sections for $^{56}$Fe, calculated
at temperatures $T=10^{10}$ K, $1.5 \cdot 10^{10}$ K and $2 \cdot
10^{10}$ K. In $^{56}$Fe the backresonances are located at an excitation
energy of $E_i \sim 7-9$ MeV. Thus they are hardly populated at these
temperatures and, as a consequence,  the backresonances influence
the cross section only at neutrino energies $E_\nu \leq 7$ MeV. Despite
the negative Q-value, the $(\nu_e,e^-)$ cross section at finite temperature
does not vanish at $E_\nu=0$. Still it remains rather small at the low
neutrino energies, even at the high temperatures.

Haxton has pointed out that, after the bounce, infalling matter might
get preheated by  $\nu_e$-absorption before it is reached by the shock
\cite{Haxton88}. This matter is expected to be mainly $^{56}$Fe. As the
average $\nu_e$ energy for this preheating process is about 15 MeV and
the matter temperature is below $T=10^{10}$ K, our calculation indicates
that the relevant cross section can be calculated neglecting temperature
effects. The respective cross sections are presented in \cite{Toivanen}.

During the core contraction $\nu_e$ absorption on nuclei is strongly
hindered by Pauli blocking of the electron in the final state. 
In step 3, we have
investigated  this blocking by introducing a blocking factor $(1-f(E_e))$
into Eq. (2), where $f(E_e)$ is a Fermi-Dirac distribution with temperature
$T$ and chemical potential $\mu_e$. We have calculated the cross section
for 3 relevant stellar conditions defined by temperature ($T$ in MeV)
and density ($\rho$ in $10^{10}$ g/cm$^3$): 
($T,\rho$)=(0.86,1.),
(1.29,10.), and (1.72,100.). The corresponding chemical potentials, 
calculated as described in \cite{Langanke00}
and assuming the same
electron-to-baryon ratios $Y_e$ as in Fig. 1 of \cite{Bruenn91} at these
densities,
are 8.3 MeV, 18.1
MeV and 36.2 MeV, respectively. The blocked cross sections are shown in
Fig. 1c. We note that, for absorption on backresonances, the electron in
the final state additionally gains the nuclear excitation energy in the
initial state and hence is less affected by the blocking. 
Obviously the importance of the backresonances is a competition of the
Boltzmann weight in the thermal ensemble, which increases with
temperature, and the Pauli blocking, which decreases the cross section.
As the chemical potential, which sets the scale for the blocking,
increases faster than the temperature the blocking becomes dominant 
with increasing temperature and density. Although the cross sections are
still finite at low neutrino energies and are clearly enhanced compared
to calculations performed at $T=0$, i.e. for the ground state only, the
respective cross section values are significantly smaller than the
competing inelastic neutrino scattering of electrons and nuclei as given
in \cite{Bruenn91}.

The nucleus $^{56}$Fe is not typical to explore finite temperature
effects on neutrino absorption cross section in supernova conditions for
3 reasons. 
(i) The matter composition is more neutronrich, when neutrino
transport is important in late stellar evolution. As a consequence, absorption
of electron neutrinos has usually positive $Q$-values and does not have to
overcome energy thresholds. (ii) Due to the Ikeda sumrule the GT
strength in the $(\nu_e, e^-)$ direction increases proportional to the
neutron excess $(N-Z)$, while the backresonance contribution is reduced
due to blocking of the neutron phase space in the final state. This
effect enlarges the relative importance of the ground state contribution
in Eq (2) with respect to the backresonances. However, there is an
interesting third point related to nuclear structure. (iii)
The energy position of
the backresonances  depends on the
pairing structure of the nucleus \cite{Langanke00}. In odd-odd nuclei
the backresonances reside at significantly lower excitation energies
($E_i \sim 1-3$ MeV) than in odd-$A$ nuclei ($E_i \sim 4-6$ MeV), while
they are largest in even-even nuclei. 
To explore the influence of these 3 effects on the finite-temperature
neutrino absorption cross sections we have performed calculations,
similar to those discussed for $^{56}$Fe in Figs. 1b and 1c, 
for the two even-even nuclei
$^{60,62}$Ni, the odd-$A$ nuclei $^{59,61}$Fe and the odd-odd nuclei
$^{60,62}$Co. 

Fig. 2 shows the temperature dependence of the absorption cross section
for these nuclei. The even-even Ni isotopes still have negative
$Q$-values and hence show a similar behavior as $^{56}$Fe. The situation
is different for the other four nuclei where neutrino absorption is
possible even for $E_\nu=0$. The influence of the Ikeda sumrule is
clearly visible if comparing the cross sections for two isotopes: The
enhancement due to finite temperature is reduced the more neutronrich
the nucleus. This effect is strengthened by the accompanying increase in 
the $Q$-value. For $^{60}$Co we observe an increase of the cross section at
finite temperature by
several orders of magnitude at low neutrino energies. This has two
reasons. As mentioned above, the backresonances reside at rather low
excitation energies ($E_i=1-2$ MeV) and are rather easily accessible by
thermal excitation. However, $^{60}$Co serves as an example for another
nuclear structure effect which is often encountered for odd-odd nuclei.
This nucleus has a ground state spin of
$J=5^+$ which introduces a strong angular momentum mismatch with the
daughter nucleus and a significantly reduced effective $Q$-value for GT
transitions. As a consequence a calculation of the absorption cross
section for odd-odd nuclei solely on the basis of the 
ground state GT distribution is not a
very accurate approximation at small neutrino energies.
For the stellar environment this inaccuracy does not matter too much as
electron blocking in the final state dominates the stellar cross
sections. This is demonstrated in Fig. 3. Although the stellar cross
sections are noticeably enhanced by about 2 orders of magnitude at low
neutrino energies due to finite temperature effects, these cross
sections are still noticeably smaller than the competing ones for inelastic
neutrino scattering on electrons and nuclei.
When compared to the stellar rates of these competing
reactions (Fig. 1 of \cite{Bruenn91}), neutrino absorption on
nuclei is important during the stellar collapse only for neutrino
energies at which finite temperature effects are negligible.

How does finite temperature affect the absorption cross section of
antineutrinos in the stellar environment? We have performed calculations,
using Eq. (2),
for the same set of nuclei as discussed above for neutrino absorption.
To understand the results shown in Fig. 4, 
one has to consider that some of the items listed
above invert for the $(\bar \nu, e^+$) reactions. Hence, i) the $Q$
values become less favorable with increasing neutron excess and the
reactions have to overcome noticeable thresholds. ii) Due to the Ikeda
sumrule the contribution of the backresonances becomes increasingly more
important with growing neutron excess. For neutronrich
nuclei the largest contribution
of the GT$_-$ strength is in the isospin $\Delta T=-1$ transition 
which reside at rather low excitation energies. 
(iii) The
energy positions of these backresonances are also dependent on the
pairing structure, again being lowest for odd-odd nuclei. Finally one has to
consider that the chemical potential for positrons is the negative of
the electron chemical potential. Thus there is no final state blocking
for antineutrino absorption under core-contraction condition.
We observe from Fig. 4 that finite-temperature effects significantly
effect the cross sections for antineutrino energies 
$E_{\bar \nu} < 10$ MeV,
leading to finite and sizable results even for $
E_{\bar \nu} < Q$. We also observe that the cross sections do not depend
too strongly on nuclear structure, i.e. they are roughly the same for
the various nuclei at a given temperature.

In summary, we have studied the effects of finite temperature on the
stellar neutrino and antineutrino absorption cross sections. We find
that the neutrino cross sections for low-energy neutrinos
are enhanced at finite temperature
mainly due to the thermal population of the backresonances. However, the
electron chemical potential in the supernova environment increases more
rapidly than the temperature. Thus the stellar absorption rates for low-energy
neutrinos are dominated by electron blocking in the final state making
the stellar neutrino absorption rate much smaller than the competing inelastic
neutrino scattering off electrons and nuclei. Neutrino absorption might
be important during the stellar collapse for higher neutrino energies
where temperature effects are unimportant. Furthermore an accurate
description of the GT distribution is probably not required at these
higher neutrino energies and the relevant rates can be obtained on the
basis of less sophisticated nuclear models, like the random phase
approxmation. These calculations, which are in progress, have to include
transitions mediated by higher multipole operators using also the
correct dependence of these operators on momentum transfer.

Antineutrino absorption cross sections on neutronrich
nuclei are strongly enhanced at finite temperature for small
antineutrino energies.   
As the stellar rates are roughly the same for different nuclei
it might be
sufficient to represent this process  by a typical nucleus in
core-collapse simulations.

We like to end with a remark on finite temperature effects on inelastic
neutrino scattering on nuclei. This process 
(together with inelastic 
scattering on electrons) 
is considered as the mechanism 
to downscatter neutrinos in energy where they can escape from the star
more easily. However, at finite temperature inelastic neutrino
scattering off nuclei will also be enhanced, mainly due to the thermal
population of the backresonances in the GT$_0$ distribution. 
These backresonance contributions will, however, increase the neutrino
energies. We are currently investigating the relevance of this effect in
more details.

\acknowledgements

Our work was supported in part 
by the Danish Research Council and
by the Swiss National Science Foundation.

\begin{figure}
\centerline{\psfig{figure=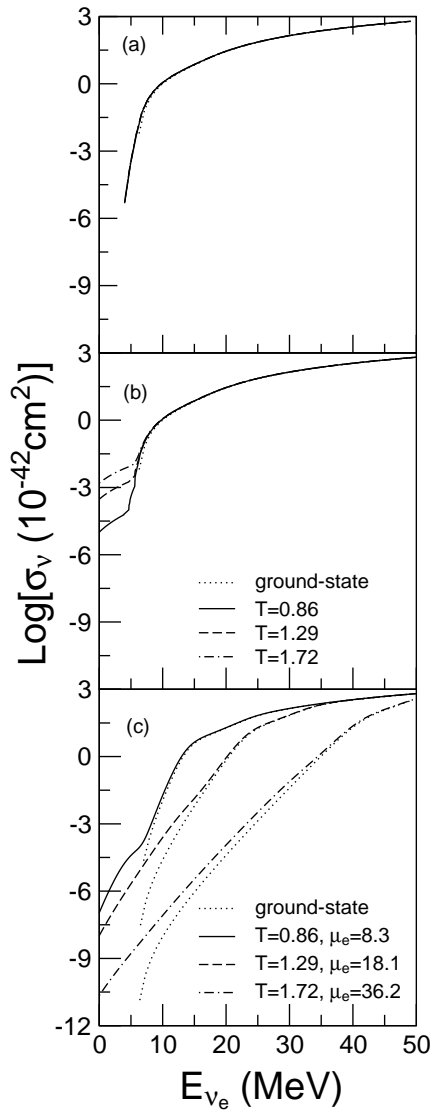,width=.40\columnwidth}}
\caption
{Absorption cross section of electron neutrinos on $^{56}$Fe. a)
Comparison of the cross section calculated for the ground state (dotted)
with the one obtained at finite temperature $T=0.86$ MeV. The calculations
have been performed using Eq. (1) where the 4 lowest
states in $^{56}$Fe have been considered in the thermal ensemble.
b) Comparison of the cross section calculated at finite temperatures
(in MeV)
with the one derived from the ground state only (dotted). The
calculations at finite temperatures have been performed using Eq. (2)
without consideration of
electron blocking in the final state;
c) the same cross sections as in b) but now considering the electron blocking
in the final state. The respective temperatures and chemical potentials
(both in MeV) are given in the figure.
}
\label{fig_fe56}
\end{figure}

\newpage

\begin{figure}
\centerline{\psfig{figure=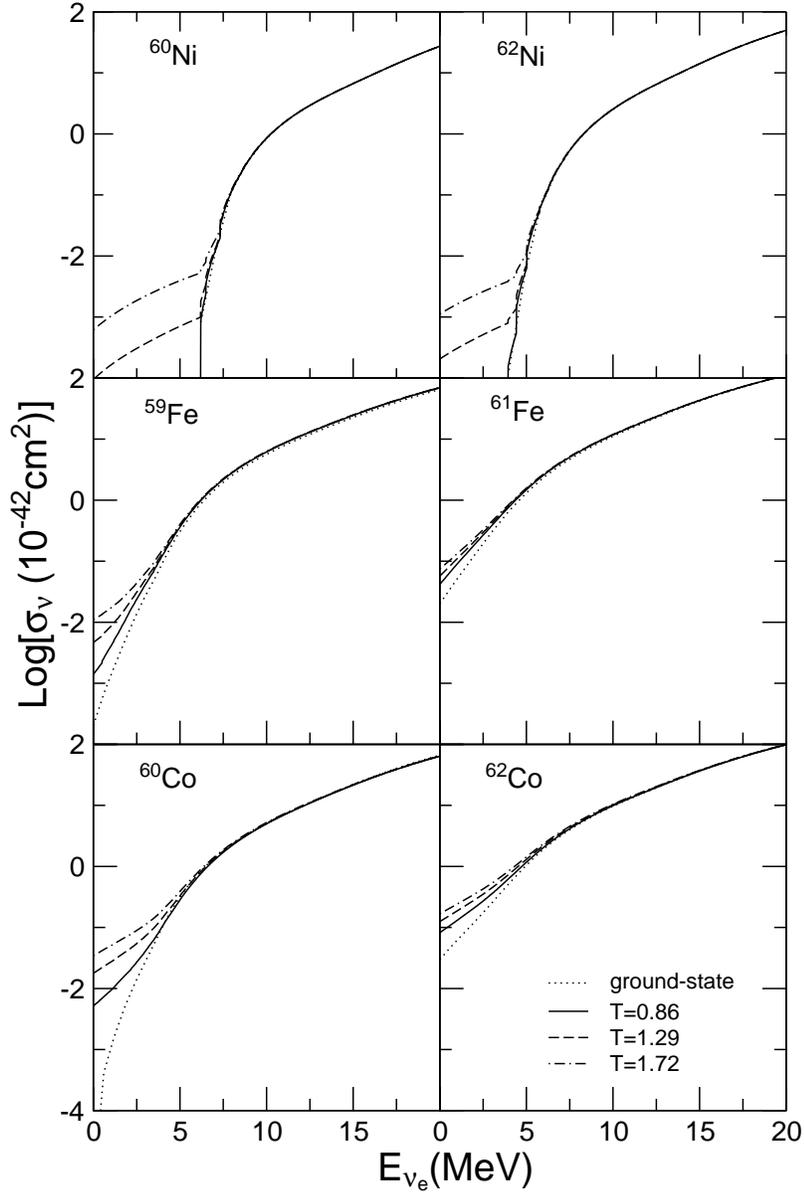,width=.8\columnwidth}}
\caption
{Absorption cross section of electron neutrinos on selected neutronrich
nuclei. The calculations have been performed for the same temperatures
as in Fig. 1b) (temperature defined in MeV in the figure) using Eq. (2).
The finite-temperature results are compared to the one derived from 
the ground state (dotted).
}
\label{Fig2}
\end{figure}

\newpage

\begin{figure}
\centerline{\psfig{figure=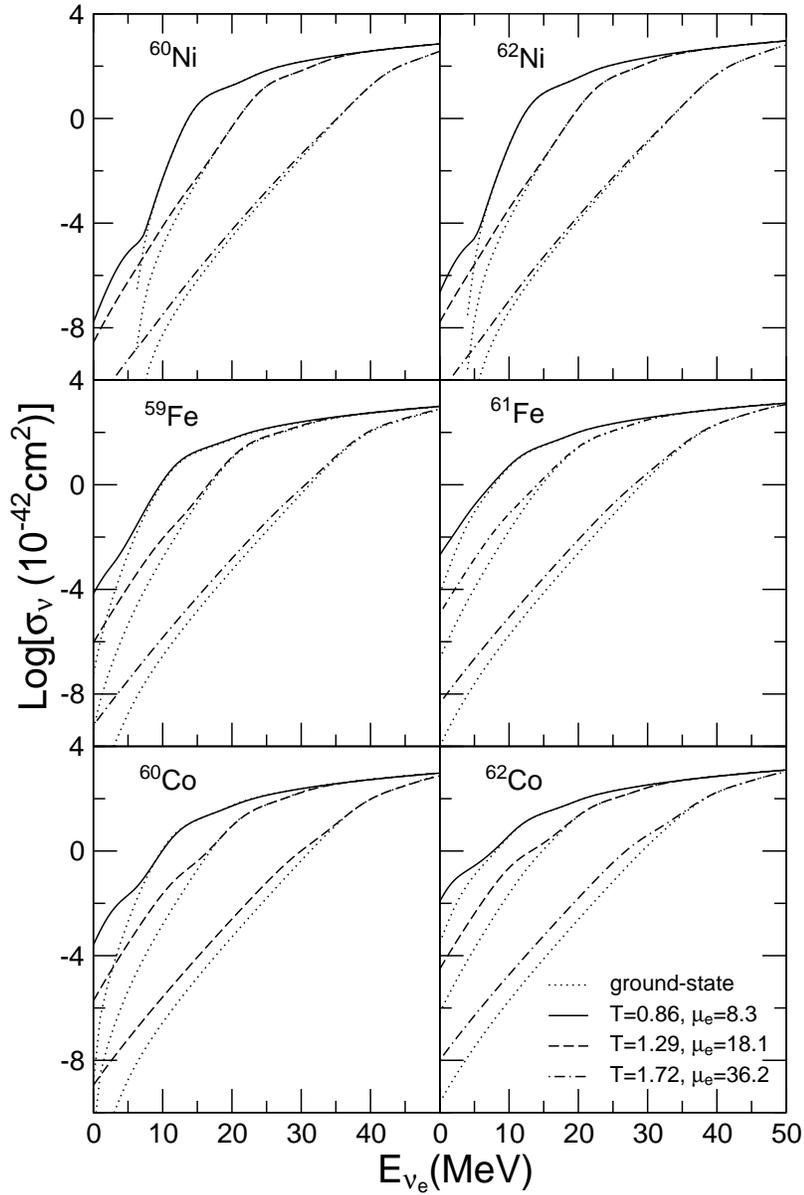,width=.8\columnwidth}}
\caption
{The same as Fig. 2, but now including the Pauli blocking of the
electron in the final state. The stellar conditions are defined by 
the temperature and the chemical potential (both in MeV). 
The finite-temperature cross sections are compared to the one derived
from the ground state alone.
}
\label{Fig3}
\end{figure}

\newpage

\begin{figure}
\centerline{\psfig{figure=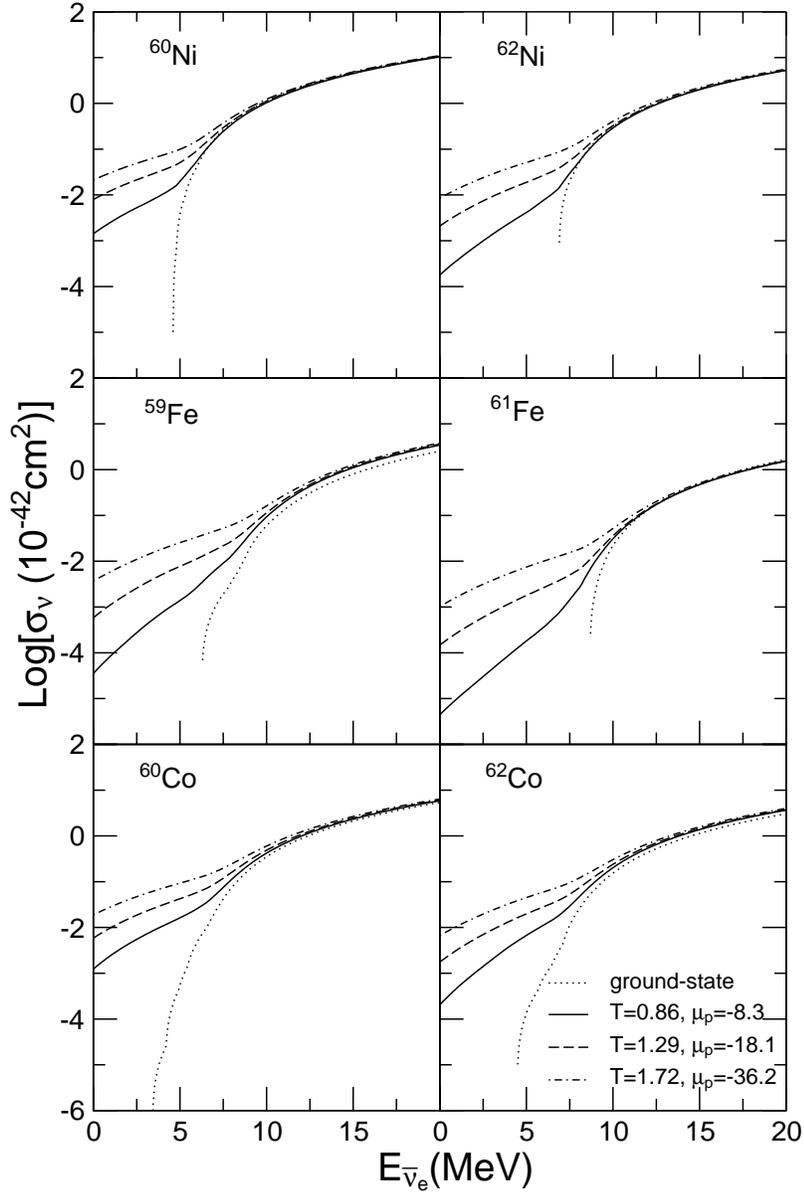,width=.8\columnwidth}}
\caption
{The same as Fig. 2, but for antineutrinos. Final state blocking for
positrons is negligible for the chosen core collapse conditions.
The backresonance contribution also includes the one for the Fermi
transition, which, however, is at too high an excitation energy to influence 
the
cross sections.
}
\label{Fig4}
\end{figure}

\end{document}